\documentclass[twocolumn]{aastex631}

\shorttitle{The Simons Observatory: the Large Aperture Telescope}
\shortauthors{The Simons Observatory Collaboration}
\graphicspath{{./}{figures/}}
\begin{document}

\title{The Simons Observatory: the Large Aperture Telescope (LAT)}

\correspondingauthor{Zhilei Xu}
\email{zhileixu@sas.upenn.edu}

\author[0000-0001-5112-2567]{Zhilei Xu}
\affiliation{Department of Physics and Astronomy, University of Pennsylvania, Philadelphia, PA, 19104 USA}
\affiliation{MIT Kavli Institute, Massachusetts Institute of Technology, Cambridge, MA, 02139 USA}

\author[0000-0002-0400-7555]{Shunsuke Adachi}
\affiliation{Kavli IPMU, WPI, UTIAS, The University of Tokyo, Kashiwa, Chiba 277-8583, Japan}

\author{Peter Ade}
\affiliation{Department of Physics and Astronomy, Cardiff University, The Parade, Cardiff CF24 3AA, UK}

\author{J. A. Beall}
\affiliation{Quantum Sensors Group, National Institute of Standards and Technology, 325 Broadway, Boulder, CO 80305, USA}

\author{Tanay Bhandarkar}
\affiliation{Department of Physics and Astronomy, University of Pennsylvania, Philadelphia, PA, 19104 USA}

\author{J. Richard Bond}
\affiliation{Canadian Institute for Theoretical Astrophysics, 60 St. George Street,  University of Toronto, Toronto, ON, M5S 3H8, Canada}

\author{Grace E.~Chesmore}
\affiliation{Department of Physics, University of Chicago, 5720 South Ellis Avenue, Chicago, IL 60637, USA}

\author[0000-0002-3266-857X]{Yuji Chinone}
\affiliation{Research Center for the Early Universe, School of Science, The University of Tokyo, Tokyo 113-0033, Japan}
\affiliation{Kavli IPMU, WPI, UTIAS, The University of Tokyo, Kashiwa, Chiba 277-8583, Japan}

\author{Steve K.~Choi}
\affiliation{Department of Physics, Cornell University, Ithaca, NY 14853, USA}
\affiliation{Department of Astronomy, Cornell University, Ithaca, NY 14853, USA}

\author{Jake A.~Connors}
\affiliation{Department of Physics, University of Colorado Boulder, Boulder, CO 80305, USA}

\author{Gabriele Coppi}
\affiliation{Department of Physics, University of Milano-Bicocca, Milano (MI), Italy}

\author{Nicholas F.~Cothard}
\affiliation{Department of Applied and Engineering Physics, Cornell University, Ithaca, NY 14853, USA}

\author{Kevin D.~Crowley}
\affiliation{Department of Physics, Princeton University, Princeton, NJ 08544, USA}

\author{Mark Devlin}
\affiliation{Department of Physics and Astronomy, University of Pennsylvania, Philadelphia, PA, 19104 USA}

\author[0000-0002-1940-4289]{Simon Dicker}
\affiliation{Department of Physics and Astronomy, University of Pennsylvania, Philadelphia, PA, 19104 USA}

\author[0000-0002-6817-0829]{Bradley Dober}
\affiliation{Department of Physics, University of Colorado Boulder, Boulder, CO 80305, USA}

\author{Shannon M.~Duff}
\affiliation{Quantum Sensors Group, National Institute of Standards and Technology, 325 Broadway, Boulder, CO 80305, USA}

\author[0000-0001-7225-6679]{Nicholas Galitzki}
\affiliation{Department of Physics, University of California San Diego, La Jolla, CA 92093, USA}

\author[0000-0001-9731-3617]{Patricio A.~Gallardo}
\affiliation{Department of Physics, Cornell University, Ithaca, NY 14853, USA}

\author[0000-0002-4421-0267]{Joseph E.~Golec}
\affiliation{Department of Physics, University of Chicago, 5720 South Ellis Avenue, Chicago, IL 60637, USA}

\author[0000-0003-1760-0355]{Jon E.~Gudmundsson}
\affiliation{The Oskar Klein Centre, Department of Physics, Stockholm University, SE-106 91 Stockholm, Sweden}

\author[0000-0001-6519-502X]{Saianeesh K.~Haridas}
\affiliation{Department of Physics and Astronomy, University of Pennsylvania, Philadelphia, PA, 19104 USA}

\author{Kathleen Harrington}
\affiliation{Department of Astronomy and Astrophysics, University of Chicago, 5640 South Ellis Avenue, Chicago, IL 60637, USA}

\author[0000-0002-4765-3426]{Carlos Hervias-Caimapo}
\affiliation{Department of Physics, Florida State University, Tallahassee, FL 32306, USA}

\author{Shuay-Pwu Patty Ho}
\affiliation{Department of Physics, Stanford University, 382 Via Pueblo, Stanford, CA 94305, USA}

\author[0000-0003-4573-4094]{Zachary B.~Huber}
\affiliation{Department of Physics, Cornell University, Ithaca, NY 14853, USA}

\author{Johannes Hubmayr}
\affiliation{Quantum Sensors Group, National Institute of Standards and Technology, 325 Broadway, Boulder, CO 80305, USA}

\author[0000-0001-7466-0317]{Jeffrey Iuliano}
\affiliation{Department of Physics and Astronomy, University of Pennsylvania, Philadelphia, PA, 19104 USA}

\author{Daisuke Kaneko}
\affiliation{Institute of Particle and Nuclear Studies, High Energy Accelerator Research Organization, 305-0801, Oho 1-1, Tsukuba, Ibaraki, Japan}

\author{Anna M.~Kofman}
\affiliation{Department of Physics and Astronomy, University of Pennsylvania, Philadelphia, PA, 19104 USA}

\author[0000-0003-0744-2808]{Brian J.~Koopman}
\affiliation{Department of Physics, Yale University, New Haven, CT 06520, USA}

\author[0000-0002-6522-6284]{Jack Lashner}
\affiliation{Department of Physics and Astronomy, University of Southern California, 3551 Trousdale Pkwy, Los Angeles, CA 90089, USA}

\author[0000-0002-5900-2698]{Michele Limon}
\affiliation{Department of Physics and Astronomy, University of Pennsylvania, Philadelphia, PA, 19104 USA}

\author{Michael J.~Link}
\affiliation{Quantum Sensors Group, National Institute of Standards and Technology, 325 Broadway, Boulder, CO 80305, USA}

\author{Tammy J.~Lucas }
\affiliation{Quantum Sensors Group, National Institute of Standards and Technology, 325 Broadway, Boulder, CO 80305, USA}

\author[0000-0003-0041-6447]{Frederick Matsuda}
\affiliation{Institute of Space and Astronautical Science (ISAS), JAXA, Sagamihara, Kanagawa 252-5210, Japan}

\author{Heather McCarrick}
\affiliation{Department of Physics, Princeton University, Princeton, NJ 08544, USA}

\author[0000-0002-8307-5088]{Federico Nati}
\affiliation{Department of Physics, University of Milano Bicocca, Piazza della Scienza, 3, Milano 20126, Italy}

\author{Michael D.~Niemack}
\affiliation{Department of Physics, Cornell University, Ithaca, NY 14853, USA}
\affiliation{Department of Astronomy, Cornell University, Ithaca, NY 14853, USA}
\affiliation{Kavli Institute at Cornell for Nanoscale Science, Cornell University, Ithaca, NY 14853, USA}

\author{John Orlowski-Scherer}
\affiliation{Department of Physics and Astronomy, University of Pennsylvania, Philadelphia, PA, 19104 USA}

\author{Lucio Piccirillo}
\affiliation{Department of Physics and Astronomy, The University of Manchester, Oxford Rd, Manchester M13 9PL, UK}

\author{Karen Perez Sarmiento}
\affiliation{Department of Physics and Astronomy, University of Pennsylvania, Philadelphia, PA, 19104 USA}

\author[0000-0002-4619-8927]{Emmanuel Schaan}
\affiliation{Lawrence Berkeley National Laboratory, One Cyclotron Road, Berkeley, CA 94720, USA}
\affiliation{Berkeley Center for Cosmological Physics, UC Berkeley, CA 94720, USA}

\author[0000-0001-7480-4341]{Maximiliano Silva-Feaver}
\affiliation{Department of Physics, University of California San Diego, La Jolla, CA 92093, USA}

\author[0000-0002-1187-9781]{Rita Sonka}
\affiliation{Department of Physics, Princeton University, Princeton, NJ 08544, USA}

\author{Shreya Sutariya}
\affiliation{Department of Physics, University of Chicago, 5720 South Ellis Avenue, Chicago, IL 60637, USA}

\author{Osamu Tajima}
\affiliation{Division of Physics and Astronomy, Kyoto University, Kitashirakawa-Oiwakecho, Sakyo-ku, Kyoto 606-8502, Japan}

\author{Grant P. Teply}
\affiliation{Department of Physics, University of California San Diego, La Jolla, CA 92093, USA}

\author{Tomoki Terasaki}
\affiliation{Department of Physics, School of Science, The University of Tokyo, Tokyo 113-0033, Japan}

\author{Robert Thornton}
\affiliation{Department of Physics and Astronomy, University of Pennsylvania, Philadelphia, PA, 19104 USA}
\affiliation{Department of Physics, West Chester University of Pennsylvania, West Chester, PA 19383, USA}

\author[0000-0002-1851-3918]{Carole Tucker}
\affiliation{Department of Physics and Astronomy, Cardiff University, The Parade, Cardiff CF24 3AA, UK}

\author{Joel Ullom}
\affiliation{Quantum Sensors Group, National Institute of Standards and Technology, 325 Broadway, Boulder, CO 80305, USA}

\author{Eve M.~Vavagiakis}
\affiliation{Department of Physics, Cornell University, Ithaca, NY 14853, USA}

\author{Michael R.~Vissers}
\affiliation{Quantum Sensors Group, National Institute of Standards and Technology, 325 Broadway, Boulder, CO 80305, USA}

\author[0000-0002-5855-4036]{Samantha Walker}
\affiliation{Quantum Sensors Group, National Institute of Standards and Technology, 325 Broadway, Boulder, CO 80305, USA}
\affiliation{Department of Astrophysical and Planetary Sciences, University of Colorado Boulder, Boulder, CO 80309, USA}

\author{Zachary Whipps}
\affiliation{Quantum Sensors Group, National Institute of Standards and Technology, 325 Broadway, Boulder, CO 80305, USA}

\author[0000-0002-7567-4451]{Edward J.~Wollack}
\affiliation{NASA/Goddard Space Flight Center, 8800 Greenbelt Rd, Greenbelt, MD 20771, USA }

\author[0000-0002-4495-571X]{Mario Zannoni}
\affiliation{Department of Physics, University of Milano Bicocca, Piazza della Scienza, 3, Milano 20126, Italy}

\author{Ningfeng Zhu}
\affiliation{Department of Physics and Astronomy, University of Pennsylvania, Philadelphia, PA, 19104 USA}

\author[0000-0001-6841-1058]{Andrea Zonca}
\affiliation{San Diego Supercomputer Center, University of California San Diego, La Jolla, California, USA}

\collaboration{60}{The Simons Observatory Collaboration}

%% Note that the \and command from previous versions of AASTeX is now
%% depreciated in this version as it is no longer necessary. AASTeX 
%% automatically takes care of all commas and "and"s between authors names.

%% AASTeX 6.31 has the new \collaboration and \nocollaboration commands to
%% provide the collaboration status of a group of authors. These commands 
%% can be used either before or after the list of corresponding authors. The
%% argument for \collaboration is the collaboration identifier. Authors are
%% encouraged to surround collaboration identifiers with ()s. The 
%% \nocollaboration command takes no argument and exists to indicate that
%% the nearby authors are not part of surrounding collaborations.

%% Mark off the abstract in the ``abstract'' environment. 
\begin{abstract}
The Simons Observatory (SO) is a Cosmic Microwave Background (CMB) experiment to observe the microwave sky in six frequency bands from 30\,GHz to 290\,GHz.
The Observatory---at $\sim$5200\,m altitude---comprises three Small Aperture Telescopes (SATs) and one Large Aperture Telescope (LAT) at the Atacama Desert, Chile.
This research note describes the design and current status of the LAT along with its future timeline. 

\end{abstract}

%% Keywords should appear after the \end{abstract} command. 
%% The AAS Journals now uses Unified Astronomy Thesaurus concepts:
%% https://astrothesaurus.org
%% You will be asked to selected these concepts during the submission process
%% but this old "keyword" functionality is maintained in case authors want
%% to include these concepts in their preprints.

%% From the front matter, we move on to the body of the paper.
%% Sections are demarcated by \section and \subsection, respectively.
%% Observe the use of the LaTeX \label
%% command after the \subsection to give a symbolic KEY to the
%% subsection for cross-referencing in a \ref command.
%% You can use LaTeX's \ref and \label commands to keep track of
%% cross-references to sections, equations, tables, and figures.
%% That way, if you change the order of any elements, LaTeX will
%% automatically renumber them.
%%
%% We recommend that authors also use the natbib \citep
%% and \citet commands to identify citations.  The citations are
%% tied to the reference list via symbolic KEYs. The KEY corresponds
%% to the KEY in the \bibitem in the reference list below. 

\section{Introduction} \label{sec:intro}

\begin{figure*}
    \centering
    \includegraphics[width=1.0\linewidth]{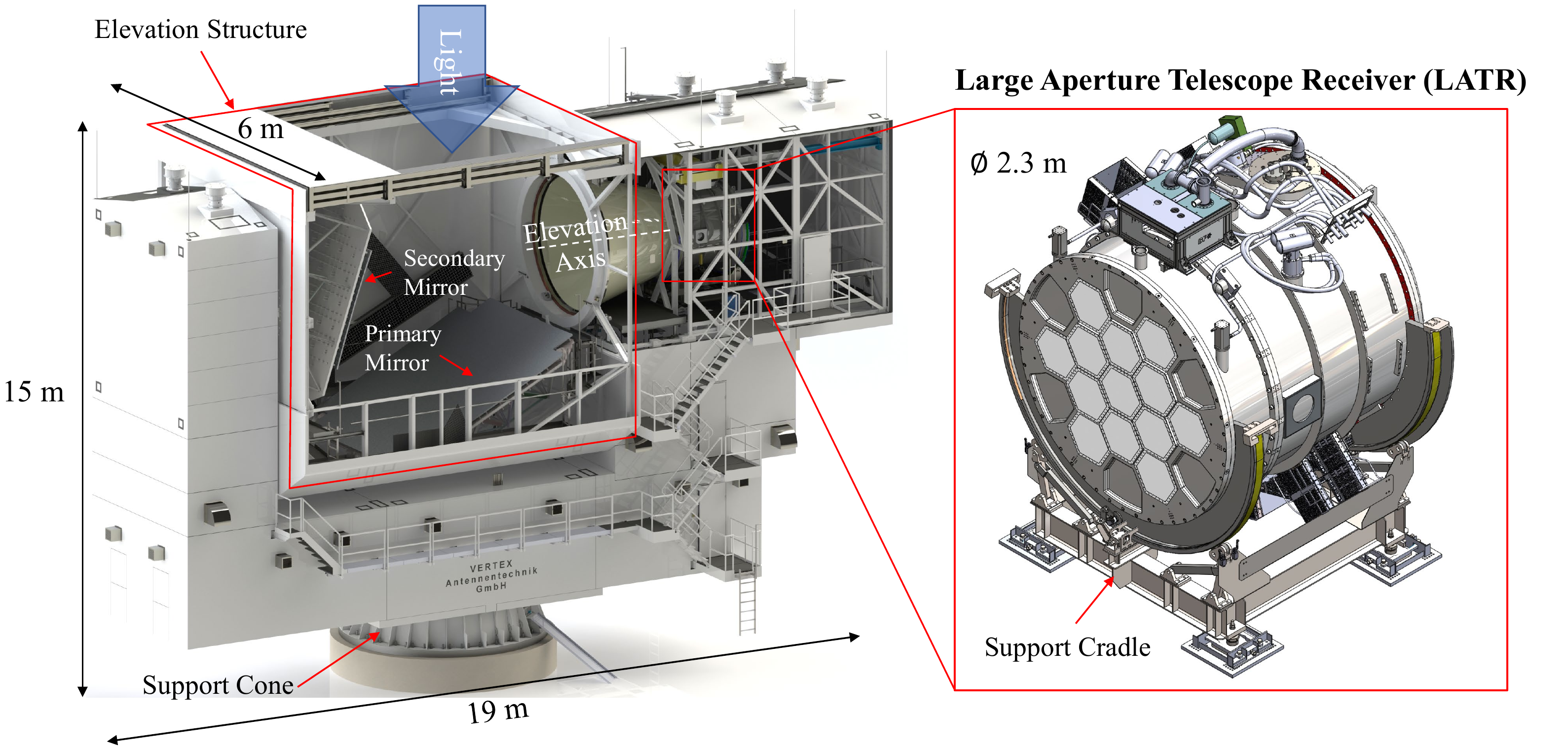}
    \caption{The LAT and its cryogenic receiver (LATR). The left image shows the rendering of the LAT. The support cone and the elevation axis are annotated, about which the telescope performs azimuth and elevation rotations during observation. The elevation structure is annotated and is looking at the zenith through the opening at the top. Within the elevation structure, two 6-m mirrors are shown to reflect light into the LATR. The right image shows the LATR from its light-receiving side. With a diameter of 2.3\,m, the LATR is the largest sub-Kelvin steerable cryogenic receiver ever built~\citep{xu/etal:2020, zhu/etal:2021}. The support cradle co-rotates the LATR with the elevation structure to mitigate optical systematics.}
    \label{fig:so_lat}
\end{figure*}

The Simons Observatory (SO)~\citep{galitzki/etal:2018, sofc19} is a cosmic microwave background experiment being built at the Chilean Atacama Desert. 
SO will have one large-aperture 6-m telescope (LAT)~\citep{parshley/etal:2018} and three small-aperture 0.5-m telescopes (SATs)~\citep{ali/etal:2020}, with a total of 60,000 polarization-sensitive transition-edge sensors (TESes)~\citep{healy/etal:2020} in the initial configuration. 

The SATs target large angular scales, mapping $\sim$10\% of the sky to a noise level of 2\,$\mu$K-arcmin in combined 90 and 150\,GHz bands.\footnote{Although two frequency bands are mentioned here, maps will be available for all six frequency bands at 30, 40, 90, 150, 230, and 290\,GHz.\label{fn:bands}} The primary science goal of the SATs is to measure the primordial perturbation tensor-to-scalar ratio ($r$), at a target level of $\sigma$($r$)=0.003. 

The LAT will map $\sim$40\% of the sky at arcminute angular resolution to an expected noise level of 6\,$\mu$K-arcmin in combined 90 and 150\,GHz bands\textsuperscript{\ref{fn:bands}} to measure the integrated mass distribution in the universe, constrain the effective number of relativistic species, measure the sum of the neutrino masses, and improve our understanding of galaxy evolution and cosmic reionization.
The LAT will also conduct a wide-field microwave survey for time-domain astronomy.
The 40\% sky coverage overlaps with future astronomical surveys, including DESI\footnote{DESI website: \url{https://www.desi.lbl.gov/}}~\citep{DESI2016} and LSST\footnote{LSST website: \url{https://www.lsst.org/}}~\citep{LSST__2019}.

\section{The LAT Design and Current Status} \label{sec:design_current_status}

The SO LAT adopts a coma-corrected, 6-m aperture, crossed-Dragone optical design~\citep{niemack:2016}. 
The telescope design delivers a 1.9-m diameter focal plane at 100\,GHz (3\,mm wavelength).
Both the 6-m mirrors are formed by rectangular panels~\citep{woody/etal:2008}: 77 panels for the primary mirror and 69 panels for the secondary mirror. 
The panels are supported by carbon fiber backup structures, and can be individually adjusted for alignment.
As shown in Figure~\ref{fig:so_lat}, the elevation structure rotates to change observation elevation.
The entire telescope structure rotates, around the support cone, to change azimuth. 

Light entering the telescope elevation structure is reflected twice before entering the telescope camera: the Large Aperture Telescope Receiver (LATR)~\citep{xu/etal:2020, zhu/etal:2021}.
The LATR is mounted on a support cradle that co-rotates along the elevation structure. The co-rotation maintains constant secondary mirror illumination for maximal optical stability.
The LATR, which measures 2.3\,m in diameter and 2.6\,m in length, contains cold optics to re-image the telescope focal plane.
The LATR contains five cryogenic temperature stages (80\,K, 40\,K, 4\,K, 1\,K, 100\,mK). 
The 80\,K, 40\,K, and 4\,K stages are cooled by pulse tube refrigerators; the 1\,K and 100\,mK stages are cooled by a dilution refrigerator.
The LATR is capable of cooling up to 13 optics tubes (OTs) with initial deployment of 7.
Each OT contains cold optics and TES detectors, which are read out by microwave-multiplexing~\citep{dober/etal:2021}. 
The LATR 100\,mK stage is capable of cooling $>$70,000 detectors (along with their support structures) to $<$100\,mK on a 1.9-m diameter focal plane.

The LAT is currently being manufactured and tested at Vertex Antennentechnik GmbH\footnote{Vertex website: \url{https://www.vertexant.com/}} in Germany. 
The SO collaboration is working closely with Vertex in the testing and validation procedure before the LAT shipment to Chile.

For the LATR, after our design was finalized, we collaborated with a vendor\footnote{Dynavac website: \url{https://dynavac.com/}} for manufacturing. The LATR vacuum and cryogenic shells were delivered to the University of Pennsylvania in 2019. 
Since then, the LATR has gone through extensive testing and integration. 
Currently, the LATR is fully equipped with the vacuum system, the thermometry system, and the cryogenic system. 
The cold optics performance and the detector/readout functionality have been validated.
Three OTs are installed in the LATR for initial testing without focal-plane modules or cold optics. 
Four more OTs are scheduled to be added soon.

In addition, the LATR has passed mechanical and cryogenic tests. The mechanical test verifies that the structure holds its $\sim$5,000\,kg weight and supports atmospheric pressure on its $\sim$27\,m$^2$ exterior surface.
The cryogenic test demonstrates that each temperature stage cools below the required base temperatures within expected time \citep{coppi/etal:2018}.
Specifically, based on the 3-OT test, the fully-equipped 13-OT 100\,mK stage is projected to reach $<$100\,mK base temperature in $<$18~days. Thermal loading on the fully-equipped 100\,mK stage is expected to be $<$70\,$\mu$W~\citep{xu/etal:2020,zhu/etal:2021}.

The scale and complexity of implementing the microwave-multiplexing technology in the LATR is unprecedented; preliminary detector/readout testing results show no signs of additional systematics compared to results from test cryostats~\citep{xu/etal:2020}.
The detector/readout system (along with the cryogenic system) is currently being tested for susceptibility to mechanical vibration, environmental temperature, external magnetic field, radio-frequency interference.

A subset of the OTs are being optically tested in a test cryostat at the University of Chicago~\citep{harrington/etal:2020}. 
These tests aims to validate the optical design of the LATR in each of the frequency bands, including system efficiency, beam shape (including far side-lobes), polarization beam, detector polarization angle, and bandpass measurement.

\section{Future Timeline} \label{sec:future_plan}

The LAT acceptance at Vertex is expected in 2021. 
Then the telescope will be shipped to the Chilean observation site, where the final assembly at the site will be conducted.
After the site assembly, final verification will be performed before Vertex hands the telescope over to the SO Collaboration. 

The LATR will have all the initial seven OTs constructed and tested in 2021, making it ready to support focal-plane modules and readout components as they become available. The LATR will be shipped to Chile in 2022 to begin commissioning. 

We expect the LAT, in the 7-OT initial configuration, to begin full scientific observation in 2023.

\begin{acknowledgements}
The Simons Observatory is supported by the Simons Foundation (Award \#457687, B.K.), the Heising-Simons Foundation, member institutions, and multiple funding agencies. Xu is supported by the Gordon and Betty Moore Foundation. Schaan is supported by the Chamberlain fellowship at LBNL. Tajima is partially supported by JSPS JP17H06134. Kaneko was supported by JSPS KAKENHI (Grant \#19K14734). Gudmundsson acknowledges grants from the Swedish Research Council (dnr. 2019-93959) and Swedish Space Agency (dnr. 139/17).
\end{acknowledgements}

%% For this sample we use BibTeX plus aasjournals.bst to generate the
%% the bibliography. The sample631.bib file was populated from ADS. To
%% get the citations to show in the compiled file do the following:
%%
%% pdflatex sample631.tex
%% bibtext sample631
%% pdflatex sample631.tex
%% pdflatex sample631.tex

\bibliography{main}{}
\bibliographystyle{aasjournal}

%% This command is needed to show the entire author+affiliation list when
%% the collaboration and author truncation commands are used.  It has to
%% go at the end of the manuscript.
%\allauthors

%% Include this line if you are using the \added, \replaced, \deleted
%% commands to see a summary list of all changes at the end of the article.
%\listofchanges

\end{document}